\documentclass[12pt]{article}

\def\be{\begin{equation}}
\def\ee{\end{equation}}
\def\bea{\begin{eqnarray}}
\def\eea{\end{eqnarray}}

\begin{document}
\title{Testing CP Conservation at KLOE}
\author{Giuseppe Mambriani and Luca Trentadue \footnote{INFN Gruppo Collegato di Parma.}
\\Dipartimento \ di \ Fisica \ \ dell' Universit\`{a} \\43100 - Parma - Italy}
\maketitle
\begin{abstract}
A proposal for maintaining CP conservation in
$K_{L}^{0}$ decays is resumed, by adding a heterodox hypothesis. One 
recovers a consistent picture with a $2\pi$ $K_{L}^{0}$ decay branching ratio 
proportional to the vertical $K_{L}^{0}$ displacement. This
can be clearly tested at the KLOE experiment, where important vertical
$K_{L}^{0}$ displacements occur.
\end{abstract}
\section{Introduction}
\qquad Bernstein, Cabibbo, and Lee \cite{Bern}, after the discovery \cite{Christ} of 2$\pi$ decay of
$K_{L}^{0}$ and the explanatory hypothesis of CP violation, considered the possibility 
that 2$\pi$ decay of $K_{L}^{0}$ 
could be due to the effect of an external field, thus saving CP
symmetry \cite{Bell}.
In this paper we try to resume the external-field hypothesis and the related
CP conservation, by introducing the heterodox assumption that the
gravitational field has a scalar component acting in opposite ways on the
$K^{0}$ and the $\overline{K^{0}}$ mixed in the$\ |K_{L}^{0}\rangle$ state. 
When suitably developed, this assumption gives a picture
which allows some simple predictions for the KLOE experiment at the
DA$\mathit{\Phi}$NE collider in Frascati. For the first time, KLOE allows
great vertical $K_{L}^{0}$\ displacements, thus permitting the
Earth's gravitational field to manifest its possible\ effects clearly. The
chief attraction of the present approach is on one hand the easiness of
testing it at KLOE, and on the other hand the simplicity introduced by CP
conservation. Some of the challenging questions opened by the repulsive gravitational
 coupling are briefly
considered.

\section{$K_{L}^{0}$ decays and a scalar component of the gravitational field}

\qquad As is well known, the $K^{0}-\overline{K^{0}}$ system has two
$\it{eigenstates}$ of the CP operator. If one assumes that CP is conserved
in all decays, the physical $K_{S}^{0}$ coincides with the
state 
$|K_{S}^{0}\rangle=$ $(|K^{0}\rangle+|\overline{K^{0}%
}\rangle)/\sqrt{2}$ having CP $\it{eigenvalue}$ $+1$. It quickly decays with
mean life $\tau_{S}$, practically always into two pions (CP
$\it{eigenvalue}$ $+1$). The other state $|K_{L}^{0}%
\rangle=(|K^{0}\rangle-|\overline{K^{0}}\rangle)/\sqrt{2}$ with CP
$\it{eigenvalue}$ $-1$, coincides with the physical $K_{L}^{0}$, 
which decays slowly (mean life $\tau_{L}\approx600$
$\tau_{S}$) along a lot of channels. Some of these channels
have final states with CP eigenvalue $+1$. This long-lived state can be
thought of as a fast mixing of $|K^{0}\rangle$ and $|\overline{K^{0}}\rangle$
performing the virtual oscillation: $K^{0}\rightarrow2\pi\rightarrow$
$\overline{K^{0}}\rightarrow2\pi\rightarrow K^{0}$, with a frequency of the
order of $1/\tau_{S}$, and a mixing energy $\hbar/\tau_{S}\approx7\times10^{-6}$ eV.

Although it's been almost forty years since the experimental discovery of
$2\pi$ decay of $K_{L}^{0}$ \cite{Christ}, the origin of the
CP violation, usually assumed to be at the root of 
$2\pi$ $K_{L}^{0}$ decay, is still judged as not fully understood. Indeed as is well
known, within the Standard Model, one can accommodate CP violation through a
parametrization of Cabibbo-Kobayashi-Maskawa's matrix (see, for instance,
\cite{DafneH} \cite{Jul} \cite{Fry}).

Bernstein, Cabibbo, and Lee \cite{Bern} a year after the discovery of $2\pi$
$K_{L}^{0}$ decay, attempted to reconcile CP conservation with it. 
They analyzed the possibility that $2\pi$ decay could
be explained in terms of the effect of an external field (they considered the three values
of the intermediate-boson spin $J$: $0$, $1$, or $2$), producing a
potential-energy difference $+V/2$ for $K^{0}$ but $-V/2$ for $\overline
{K^{0}}$. By means of a relativistic time evolution equation, they found for
the complex $2\pi$ $K_{L}^{0}$ branching ratio $\epsilon$,
when $V \ll \hbar/\tau_{S}$:
\begin{equation}
\epsilon\approx\frac{\gamma V(\gamma)}{2}\left[  
\left(  m_{L}-m_{S}\right)  c^{2}-\frac{i\hbar}{2} \left(  \frac{1}{\tau_{S}}-\frac{1}{\tau_{L}%
}\right) \right]  ^{-1/2},\label{EpsC}%
\end{equation}
where $\gamma$ is the kaon Lorentz factor in the laboratory frame, and
$m_{L}$ and $m_{S}$ are the masses of
$K_{L}^{0}$ and $K_{S}^{0}$, respectively.
Starting from general assumptions for the classical Lagrangian, the intrinsic
$V$ dependence on $\gamma$ must be expected to be: $V(\gamma
)=V_{0}$ $\gamma^{J-1}$ \cite{Thirr}, where
$V_{0}$ is the potential energy seen by the
particle when at rest, relative to the field source. In Ref.\cite{Bern}, the
assumption that $V_{0}$ is the fourth component of an unknown
vector field ($J=1$), has been made, then, from (\ref{EpsC}), it follows that
$|\epsilon|$ must be proportional to $\gamma$. After a few years, the
experiments \cite{K-66}\cite{Fitch}\cite{K-72}\ showed that $|\epsilon|$ was
independent of $\gamma$. Therefore the external-field CP-conserving
interpretation of $2\pi$ $K_{L}^{0}$\ decay was dismissed.

Actually, if in (\ref{EpsC}) one considers a scalar field ($J=0$), one has
$\left|  \epsilon\right|  $ independent of $\gamma$, as the experiments
require. The questions arise whether this field can be identified with a scalar 
component of the Earth's
gravitational field, as well as through this scalar component, the Earth can exert
an antigravitational force on $\overline{K^{0}}$. These\ questions raise a web
of problems, which will be only briefly\ considered below.

By assuming that the Earth's gravitational field has a scalar component, and
by identifying $V_{0}/2$ with $m_{K}g_{E}\Delta\zeta$ \cite{Good}, whith $m_{K}$
the kaon mass, $g_{E}$ the Earth fall acceleration, and
$\Delta\zeta$ the vertical $K_{L}^{0}$ displacement, one
finds:

\begin{equation}
|\epsilon|\approx m_{K}g_{E}\Delta\zeta \left [
\left(  m_{L}-m_{S} \right ) ^{2} c^{4}+\frac{{\hbar}^{2}}{4} \left (  
\frac{1}{\tau_{S}}-\frac{1}{\tau_{L}}\right )^{2} \right ]  
^{-1/2}=m_{K} g_{E}\;\Delta\zeta \Lambda
,\label{Meps}%
\end{equation}
where $\Lambda=(1.231\pm0.002)\times10^{24}$ J$^{-1}$, as follows from using
standard values \cite{PDG} for kaon properties.

For KLOE, $\Delta\zeta$\ is the vertical displacement between the small
intersection region of DA$\mathit{\Phi}$NE $e^{-}$ and $e^{+}$\ beams (within
which $\mathit{\Phi}$ decays into kaons) and the $K_{L}^{0}$
decay vertex. The maximum effective $\Delta\zeta$ is roughly 1.5 m, for which
(\ref{Meps}) gives\ $|\epsilon|=|\eta_{+-}|\approx15\times10^{-3}$, that is,
roughly seven times the standard value ($|\eta_{+-}|=(2.27\pm0.02)\times
10^{-3}$ \cite{PDG}.\ According to this approach, the first effect that KLOE
should have to observe, is an average $|\eta_{+-}|$\ value
three to four times greater than the standard one. This large average
$|\eta_{+-}|$ could be observed with a number of $K_{L}^{0}$
much smaller than the number forecast for KLOE design targets.

For instance, when collecting a sample of 10$^{4}$ $K_{L}^{0}$
with a $\Delta\zeta$ within $0.95$ m and $1.05$ m 
(inside the horizontal slabs
below and above the production point), one finds one hundred of
$K_{L}^{0}$ decaying into two charged pions,\ that is,\ a
branching ratio $|\eta_{+-}|=(10\pm1)\times10^{-3}$. Such a value would
practically exclude the standard picture. Better information can of course be
obtained by collecting \textit{e.g.} $10^{6}$ $K_{L}^{0}$
inside the whole KLOE volume. By subdividing them, for instance, in six bins
of $\Delta\zeta$ between 0.3 m and 1.5 m, one may look for the possible
proportionality relationship between $|\eta_{+-}|$\ and $\Delta\zeta$, and should
it be found, by comparing the experimental slope with $\Lambda
m_{K}g_{E}=(1070\pm2)\times10^{-5}$ m$^{-1}$,
where $m_{K}=$ $497.672\pm0.031$ MeV \cite{PDG} and
$g_{E}=9.8$ ms$^{-2}$.

Moreover, one must find small values of $|\eta_{+-}|$ at low $\Delta\zeta$.
For instance, once $10^{6}$ $K_{L}^{0}$ are collected, 
roughly $7.5\times10^{4}$ $K_{L}^{0}$ decays 
along all channels, will be observed inside a horizontal slab $20$ cm 
thick placed from $10$ cm
below to $10$ cm above the production point. With the standard $|\eta_{+-}|$
value one must expect $170\pm13$ decays into two charged pions, while
following the present approach one must detect only $40\pm7$ decays, with a
difference of nearly $9$ standard deviations.

Most information on the $K^{0}-\overline{K^{0}}$ system, has been obtained
from horiz- ontal-beam experiments, to which of course the present gravitational
interpretation must apply as well. When considering a nearly horizontal
$K_{L}^{0}$\ beam, $V_{0}$ can be due to the
gravitational fields of the Earth and of the Sun. At the Earth's surface, the
latter is much smaller than the former. For the Earth's field the
vertical effective displacements are rather small since they are mainly linked
to the $K_{L}^{0}$ beam divergency. For the Sun's field the
displacements projected along the field direction, can also be a few thousand
times greater when considering long $K_{L}^{0}$\ beams.

The papers reporting experimental $\left|  \epsilon\right|  $ values, give in
general little information, if any, concerning beam geometry. However, the
first relevant information on decay amplitudes, comes from the experiments
\cite{Christ}\cite{K-66}\cite{Fitch}\cite{K-72}\cite{K--79}, where
the $K_{L}^0$ beam lenght was markedly shorter than $100$ m. Thus, the unique
possibility of having $V_{0}$ different from zero, is linked to
a possible effect of the Earth's field, when a vertical displacement due to a
small vertical component of $K_{L}^{0}$ velocity, occurs. A
first set of experiments up to 1972 \cite{Christ}\cite{K-66}\cite{Fitch}%
\cite{K-72}, reported $\left|  \epsilon\right|  $ values grouped around
$1.95\times10^{-3}$, and all have roughly $\Delta\zeta\approx0.2$ m, as far as
is possible to judge from the beam size in the decay region. By taking
$\Delta\zeta$=$(0.20\pm0.04)$ m (a somewhat arbitrary $20\%$ error has been
assumed), and $\left|  \epsilon\right|  =(1.95\pm0.2)\times10^{-3}$ as
correlated values, eq.(\ref{Meps}) gives:

$\;\;\;\;\;\;\;\;\;\;\;\;\;\;\;\;\;\;\;\;\;g_{E}=\left|  \epsilon \right|  /(\Lambda m_{K} \Delta \varsigma)= 8.9\pm 2.7 $ m {s}$^{-2}$,

\vskip 0.3cm 
\noindent which is consistent with the standard\textit{\ }$g_{E}%
$\textit{\ }value. Since 1973 higher values of $\left|  \epsilon\right|  $
have been reported \cite{K--79} grouped finally around $2.27\times10^{-3}$
(the presently accepted value), perhaps with slightly larger $\Delta\zeta$ values.

If the present gravitational interpretation applies, the effect of the Earth's
field can introduce important biases in an experiment such as CPLEAR at CERN,
where the effective maximum vertical displacement is roughly 0.4 m. This would
make it necessary to reanalyze all CPLEAR results, such as those concerning CP
violation parameters and the so-called direct T-reversal violation.

The four experiments NA31 and NA48 at CERN, and E731 and
KTeV at Fermilab, have utilized and utilize long $K_{L}^{0}$ 
horizontal beams. Thus, their results could be biased by the effects of both the
Earth's and the Sun's field. The latter effect depends on the beam
length projected along the direction of the solar field, and this projected length
continuously varies owing to Earth rotation and revolution. Thus, these
long-beam experiments could be affected by non-trivial systematic errors, and
their results, such as those concerning 
$Re(\epsilon^{\prime}/\epsilon)$ and time reversal $T$ violation, would all have to be reanalyzed.

Among the various topics concerning the $K^{0}$-$\overline{K^{0}}$ system, let
us consider an argument due to Sakurai and Wattemberg \cite{Sakur}. They
elegantly argued that CP violation in $2\pi$ $K_{L}^{0}$ decay
is conclusively demonstrated by so-called soft regeneration, first observed by
Fitch \textit{et al}. \cite{Fitch}. This argument does not apply if one
assumes that $K^{0}$ and $\overline{K^{0}}$ have opposite gravitational
behavior, with the possibility that CP is an exact symmetry \cite{AntiS}.

Coherently with the starting hypothesis of antigravity, one must expect
that, for instance, antiprotons, antineutrons, and antiatoms are gravitationally repelled
from the Earth, thus \textit{antifalling} with the acceleration modulus
$g_{E}$. One must take into account that antigravity,
besides introducing a lot of problems in the standard physical picture, cannot
coexist with the equivalence principle \cite{Eq-Prin}. Actually, it seems
possible to maintain the agreement with a very large part of the known
phenomena, by embedding the antigravity assumption in a set of suitable
hypotheses \cite{Mamb}. In this way, it seems possible to avoid all paradoxes
and difficulties (such as violation of energy conservation, causality
violation, possible CPT violations, etc.), which antigravity implies
when directly inserted into the standard physical picture, It also seems
possible to circumvent any conflict with the so-well tested proportionality
between the inertial and the gravitational mass \cite{MiMg}.

\section{Concluding remarks}

\qquad For the first time the KLOE experiment at DA$\mathit{\Phi}$NE offers
the possibility of having important vertical $K_{L}^{0}$ displacements, thus allowing the Earth's gravitational field to
manifest its possible effects clearly. The proposal advanced in 1964 by Bernstein,
Cabibbo, and Lee has been resumed, by adding
to it the heterodox hypothesis that the gravitational field has a
scalar component acting in opposite ways on the $K^{0}$ and $\overline{K^{0}}$
mixed in $|K_{L}^{0}\rangle$. We have shown that the 
CP-conserving mechanism seems to fit the known $K_{L}^{0}$ properties well. 
The main new consequence is the proportionality between the modulus of $2\pi$
$K_{L}^{0}$ decay branching ratio and the vertical
$K_{L}^{0}$ displacement. At KLOE, this proportionality should
be observed with a relatively small number of $2\pi$ $K_{L}^{0}$ decays.
If KLOE data should validate this proportionality,  it would become necessary 
to consider the above sketched questions, and many other related topics and problems as well.


\begin{thebibliography}{99}
\bibitem {AntiS}In the so-called soft regeneration experiments \cite{Fitch},
low-density regenerators are used, so that the modulus of the complex
regeneration parameter $\rho$ is roughly equal to $\left|  \epsilon\right|
$.\ The experiments show that there is interference between $\rho$ and
$\epsilon$ and that, therefore, the final two-pion states of
$K_{S}^{0}$ and $K_{L}^{0}$ decays, are
identical. This identity provides an absolute criterion for distinguishing
matter from antimatter, because in an experiment performed on an Antiearth the
relative $\rho$ and $\epsilon$ sign would change, thus exhibiting a
matter-antimatter asymmetry, that is, a clear signature of CP violation. The
Sakurai-Wattemberg argument does not apply if one assumes that $K^{0}$ and
$\overline{K^{0}}$ have opposite gravitational behavior. In fact, when
observing $2\pi\ K_{L}^{0}$ decay on an Antiearth, not only
$\rho$ would reverse its sign as in the standard physical picture, but also
$\epsilon$ would do the same, owing to the reversed gravitational interactions
of both $K^{0}$ and $\overline{K^{0}}$. Then, the relative sign of $\rho$ and
$\epsilon$ would not change, showing complete symmetry between matter and
antimatter, thus reopening the possibility that CP is an exact symmetry.
\end{thebibliography}
\end{document}